\documentclass{PoS}
\usepackage{bm}

\title{Excited-state hadron masses using the stochastic LapH method}

\ShortTitle{Excited-state hadron masses using the stochastic LapH method}

\author{John Bulava$^a$, Justin Foley$^b$, You-Cyuan Jhang$^c$, 
        Keisuke~J.~Juge$^d$, David~Lenkner$^c$,
        \speaker{Colin Morningstar}$^c$, and Chik Him~Wong$^e$\\
\llap{$^a$} CERN, Physics Dept., 1211 Geneva 23, Switzerland\\
\llap{$^b$} Dept.~of Physics and Astronomy, University of Utah, 
        Salt Lake City, UT 84112, USA\\
\llap{$^c$} Dept.~of Physics, Carnegie Mellon University, 
        Pittsburgh, PA 15213, USA\\
\llap{$^d$} Dept.~of Physics, University of the Pacific, 
        Stockton, CA 95211, USA\\
\llap{$^e$} Dept.~of Physics, University of California San Diego,
        La Jolla, CA 92093, USA}

\abstract{Progress in computing the spectrum of excited baryons and mesons 
in lattice QCD is described. Large sets of spatially-extended hadron 
operators are used.  The need for multi-hadron operators in addition
to single-hadron operators is emphasized,
necessitating the use of a new stochastic method of treating the
low-lying modes of quark propagation which exploits Laplacian Heaviside
quark-field smearing.  A new glueball operator is tested and 
computing the mixing of this glueball operator with a quark-antiquark
operator and multiple two-pion operators is shown to be feasible.  Some of
our initial results show warning signs about extracting high-lying
resonance energies using only single-hadron operators.}

\FullConference{ The XXIX International Symposium on Lattice Field 
Theory - Lattice 2011\\ July 10-16, 2011\\
Squaw Valley, Lake Tahoe, California}

\begin{document}

\section{Introduction}
In a series of papers\cite{baryons2005A,baryons2005B,baryon2007,nucleon2009,
Bulava:2010yg,StochasicLaph}, we have been striving to 
compute the finite-volume stationary-state energies of QCD using Markov-chain
Monte Carlo integration of the QCD path integrals formulated on a
space-time lattice. Such calculations are very challenging.  
The use of carefully designed quantum field operators is crucial 
for accurate determinations of low-lying energies. To study a particular state 
of interest, the energies of all states lying below that state must first be 
extracted, and as the pion gets lighter in lattice QCD simulations, more and 
more multi-hadron states lie below the masses of the excited resonances.  The 
evaluation of correlations involving multi-hadron operators contains new 
challenges since not only must initial to final time quark propagation be 
included, but also final to final time quark propagation for a large number
of times must be incorporated.  The masses and widths of resonances must be 
deduced from the discrete spectrum of finite-volume stationary 
states for a range of box sizes.

To compute the QCD stationary-state energies, the matrices $C_{ij}(t)$ of 
temporal correlations of sets of single-hadron and multi-hadron operators are
estimated using the Monte Carlo method.  For an $N\times N$ matrix, the
$N$ eigenvalues of $C(t_0)^{-1/2}C(t)C(t_0)^{-1/2}$ tend to $\exp(-E_k (t-t_0))$
for large $t$ and fixed $t_0$, where the decay rates $E_k$ are the $N$ 
lowest-lying stationary-state energies that can be produced from the vacuum
by the operators used.  To compute the correlations involving isoscalar
mesons and good multi-hadron operators, quark propagation from all spatial
sites on one time slice to all spatial sites on another time slice 
are needed, and propagation from the sink time to the sink time are needed
for a large number of sink times.  A new method known as the stochastic
LapH method\cite{StochasicLaph} has been introduced to make such
computations accurate and practical in large volumes.

\section{The Stochastic LapH method}

Finding better ways to stochastically estimate slice-to-slice quark propagators
is crucial to the success of our excited-state hadron spectrum project
at lighter pion masses.  We have developed and tested a new scheme 
which combines a new way of smearing the quark field with a new way of 
introducing noise and dilution projectors. The new quark-field smearing scheme, 
here called Laplacian Heaviside (LapH), has been described in 
Ref.~\cite{distillation2009} and is defined by
\begin{equation}
\widetilde{\psi}(x) = 
 \Theta\left(\sigma_s^2+\widetilde{\Delta}\right)\psi(x),
\end{equation}
where $\widetilde{\Delta}$ is the three-dimensional covariant Laplacian
in terms of the stout-smeared gauge field and $\sigma_s$ is the smearing
parameter.  The gauge-covariant Laplacian operator is
ideal for smearing the quark field since it is one of the simplest operators
that locally averages the field in such a way that all relevant symmetry
transformation properties of the original field are preserved.  
Let $V_\Delta$ denote the unitary matrix whose columns are the eigenvectors
of $\widetilde{\Delta}$, and let $\Lambda_\Delta$ denote a diagonal matrix
whose elements are the eigenvalues of $\widetilde{\Delta}$ such that
$
    \widetilde{\Delta}=V_\Delta\ \Lambda_\Delta\ V_\Delta^\dagger.
$
The LapH smearing matrix is then given by
$
    {\cal S} = V_\Delta\ \Theta\left(\sigma_s^2+\Lambda_\Delta\right)
   \ V_\Delta^\dagger. 
$
Let $V_s$ denote the matrix whose columns are in one-to-one correspondence
with the eigenvectors associated with the $N_v$ lowest-lying eigenvalues of 
$-\widetilde{\Delta}$ on each time slice.  Then our LapH smearing matrix 
is well approximated by the Hermitian matrix
$
   {\cal S}=V_s\ V_s^\dagger.
$
Evaluating the temporal correlations of our hadron operators
requires combining Dirac matrix elements associated with various quark 
lines ${\cal Q}$.  Since we construct our hadron operators out of
covariantly-displaced, smeared quark fields, each and every quark line 
involves the following product of matrices:
\begin{equation}
 {\cal Q} = D^{(j)}{\cal S}M^{-1}{\cal S}D^{(k)\dagger},
\end{equation}
where $D^{(i)}$ is a gauge-covariant displacement of type $i$. 

An exact treatment of such a quark line is very costly and wasteful. Given our
use of the Monte Carlo method to evaluate the path integrals
over the gauge link variables, the statistical errors in our estimates of
the hadron correlators are ultimately limited by the statistical fluctuations
arising from the gauge-field sampling.  Thus, we only need to estimate the
quark lines to an accuracy comparable to the gauge noise from the Monte
Carlo method.  Such estimates can be obtained with far fewer inversions
than required by an exact treatment of the quark lines.

Random noise vectors $\eta$ whose expectations satisfy
$E(\eta_i)=0$ and $E(\eta_i\eta_j^\ast)=\delta_{ij}$ are useful for 
stochastically estimating the inverse of a large matrix $M$ 
as follows.  Assume that for each of $N_R$ 
noise vectors, we can solve the following
linear system of equations: $M X^{(r)}=\eta^{(r)}$ for $X^{(r)}$.
Then $X^{(r)}=M^{-1}\eta^{(r)}$, and $E( X_i \eta_j^\ast ) = M^{-1}_{ij}$
so that a Monte Carlo estimate of $M_{ij}^{-1}$ is given by
$
  M_{ij}^{-1} \approx \lim_{N_R\rightarrow\infty}\frac{1}{N_R}
 \sum_{r=1}^{N_R} X_i^{(r)}\eta_j^{(r)\ast}.
$
Unfortunately, this equation usually produces stochastic estimates with 
variances which are much too large to be useful.  Variance reduction is
done by \textit{diluting} the noise vectors.
A given dilution scheme can be viewed as the application of a complete
set of projection operators $P^{(a)}$. Define
$
  \eta^{[a]}_k=P^{(a)}_{kk^\prime}\eta_{k^\prime} ,
$
and further define $X^{[a]}$ as the solution of
$
   M_{ik}X^{[a]}_k=\eta^{[a]}_i,
$
then we have
\begin{equation}
   M_{ij}^{-1}\approx \lim_{N_R\rightarrow\infty}\frac{1}{N_R}
 \sum_{r=1}^{N_R} \sum_a X^{(r)[a]}_i\eta^{(r)[a]\ast}_j.
\label{eq:diluted}
\end{equation}
The use of $Z_4$ noise ensures
zero variance in the diagonal elements $E(\eta_i\eta_i^\ast)$.

The effectiveness of the variance reduction depends on the
projectors chosen.  With LapH smearing, noise vectors 
$\rho$ can be introduced \textit{only in the LapH subspace}.  The noise
vectors $\rho$ now have spin, time, and Laplacian eigenmode number
as their indices.  Color and space indices get replaced by Laplacian
eigenmode number.  Again, each component of $\rho$ is a random
$Z_4$ variable so that $E(\rho)=0$ and $E(\rho\rho^\dagger)=I_d$.
Dilution projectors $P^{(b)}$ are now matrices in the LapH subspace. 
In the stochastic LapH method, a quark line on a gauge configuration
is evaluated as follows:
\begin{eqnarray}
 {\cal Q}  &=&   D^{(j)} {\cal S} M^{-1} {\cal S} D^{(k)\dagger},\nonumber\\
  &=&   D^{(j)} {\cal S} M^{-1} V_s V_s^\dagger D^{(k)\dagger},\nonumber\\
  &=& \textstyle\sum_b  D^{(j)} {\cal S} M^{-1} V_s P^{(b)}P^{(b)\dagger} 
 V_s^\dagger D^{(k)\dagger},\nonumber\\
&=& \textstyle\sum_b D^{(j)} {\cal S} M^{-1} V_sP^{(b)}E(\rho\rho^\dagger)
  P^{(b)\dagger}  V_s^\dagger D^{(k)\dagger},\nonumber\\
 &=& \textstyle\sum_b E\Bigl( \! D^{(j)} {\cal S} M^{-1} V_sP^{(b)}\rho
     \, (D^{(k)} V_s P^{(b)}  \rho)^\dagger \!\Bigr).
\end{eqnarray}
For a noise vector labelled by index $r$, displaced-smeared-diluted quark source 
and quark sink vectors can be defined by
\begin{eqnarray}
 \varrho^{(r)[b](j)} &=&  D^{(j)} V_s P^{(b)}\rho^{(r)},\\
 \varphi^{(r)[b](j)} &=& D^{(j)} {\cal S} M^{-1}\ V_s P^{(b)}\rho^{(r)}, 
\end{eqnarray}
and each quark line on a given gauge configuration can be estimated using
\begin{equation}
 {\cal Q}_{uv} \approx \frac{1}{N_R}\sum_{r=1}^{N_R}\sum_b  \varphi^{(r)[b](j)}_u
  \  \varrho^{(r)[b](k)\ast}_v,
\end{equation}
where the subscripts $u,v$ are compound indices combining space, time, color,
and spin.

Our dilution projectors are
products of time dilution, spin dilution, and Laph eigenvector dilution
projectors.  For each type (time, spin, Laph eigenvector) of dilution, we
studied four different dilution schemes.  Let $N$ denote the dimension
of the space of the dilution type of interest.  For time dilution, $N=N_t$
is the number of time slices on the lattice.  For spin dilution, $N=4$ is
the number of Dirac spin components.  For Laph eigenvector dilution, $N=N_v$
is the number of eigenvectors retained.  The four schemes we studied
are defined below:
\[ \begin{array}{lll}
P^{(a)}_{ij} = \delta_{ij},              & a=0,& \mbox{(no dilution)} \\
P^{(a)}_{ij} = \delta_{ij}\ \delta_{ai},  & a=0,\dots,N-1 & \mbox{(full dilution)}\\
P^{(a)}_{ij} = \delta_{ij}\ \delta_{a,\, \lfloor Ki/N\rfloor}& a=0,\dots,K-1, & \mbox{(block-$K$)}\\
P^{(a)}_{ij} = \delta_{ij}\ \delta_{a,\, i\bmod K} & a=0,\dots,K-1, & \mbox{(interlace-$K$)}
\end{array}\]
where $i,j=0,\dots,N-1$, and we assume $N/K$ is an integer.  We use a triplet
(T, S, L) to specify a given dilution scheme, where ``T'' denotes time,
``S'' denotes spin, and ``L'' denotes Laph eigenvector dilution.  The schemes
are denoted by 1 for no dilution, F for full dilution, and B$K$ and I$K$ for
block-$K$ and interlace-$K$, respectively.  For example, full time and spin
dilution with interlace-8 Laph eigenvector dilution is denoted by
(TF, SF, LI8).  Introducing diluted noise in this
way produces correlation functions with significantly reduced variances,
yielding nearly an order of magnitude reduction in the statistical error
over previous methods.  The volume dependence of this new method was found 
to be very mild, allowing the method to be useful on large lattices.
For all forward-time quark lines, we use dilution scheme (TF, SF, LI8),
and for all same-sink-time quark lines, we use (TI16, SF, LI8).
Interlacing in time makes the calculation of a large number of sink-to-sink
diagrams possible with a feasibly small number of Dirac matrix inversions.
The use of dilution projectors that interlace in time is perhaps the most
important factor behind the success of the stochastic LapH method.

Details on how the temporal correlations of hadron operators are evaluated are
given in Ref.~\cite{StochasicLaph}.  A very useful feature of the method is the 
fact that the hadron correlators completely factorize into a function associated
with the sink time slice $t_F$, and another function associated with the source time 
slice $t_0$.  Summations over color, spin, and spatial sites at the
source can be completely separated from the color, spin, and spatial
summations at the sink.  The stochastic LapH method leads to complete 
factorization of hadron sources and sinks in temporal correlations, which greatly 
simplifies the logistics of evaluating correlation matrices involving large numbers
of operators. Implementing the Wick contractions of the quark lines is also 
straightforward.  Contributions 
from different Wick orderings within a class of quark-line diagrams differ
only by permutations of the noises at either the source or the sink.

\section{First results}
\begin{figure}[t]
  \includegraphics[width=6in,bb=0 15 567 299]{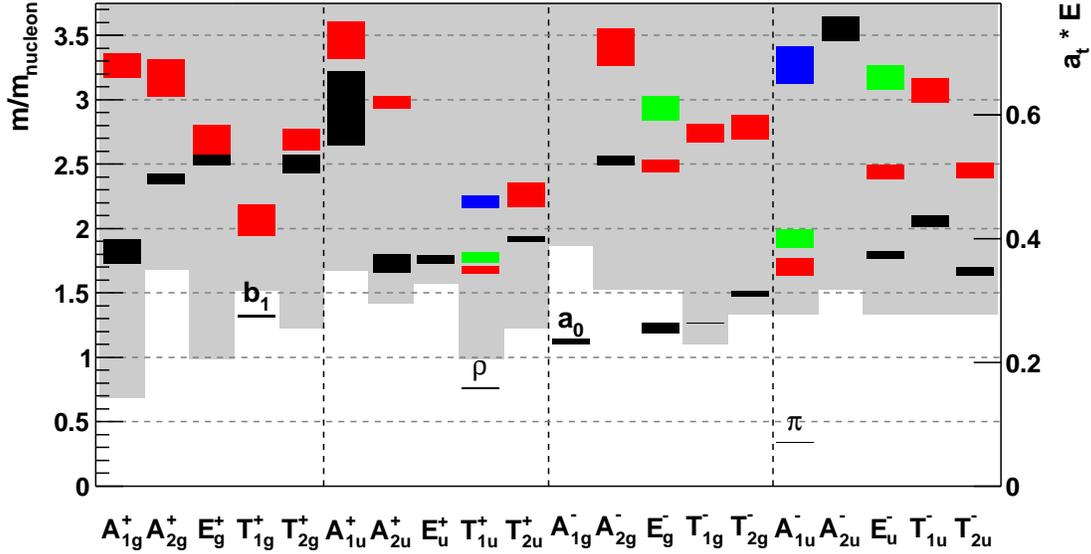}
  \caption{Masses of the isovector mesons in terms of the nucleon mass
using the stochastic LapH method with 170 gauge configurations on a $24^3\times 128$
anisotropic lattice.  The pion mass is about $m_\pi\sim 390$~MeV.
In the irrep labels, the letters with numerical subscripts refer
to the point group $O_h$ irreps, the subscripts  $g$ and $u$ refer to even and 
odd parity, respectively, and the superscripts
$\pm$ refer to $G$-parity.  Only single-hadron operators were used with dilution
scheme $(TF,SF,LI8)$.  The shaded region indicates the threshold locations for
multi-hadron energy levels.  We emphasize that extractions of energies in the
shaded regions could be complicated by ``false plateaux'' unless multi-hadron
operators are used. \label{fig:isovectorspec}}
\end{figure}

Our first results for the isovector mass spectrum on a large $24^3\times 128$
anisotropic lattice are shown in Fig.~\ref{fig:isovectorspec}.  The pion mass
is about $m_\pi\sim 390$~MeV here.  These results are not finalized since
only single-hadron operators were used.  The threshold locations for multi-hadron
energy levels, assuming the interaction energies are small, are indicated by the
shaded region.  Extractions of energies in the
shaded regions can only be considered reliable if multi-hadron operators are used,
in addition to the single-hadron operators. The inclusion of the multi-hadron
operators is in progress.

\begin{figure}[t]
  \includegraphics[width=6in,bb=0 15 567 231]{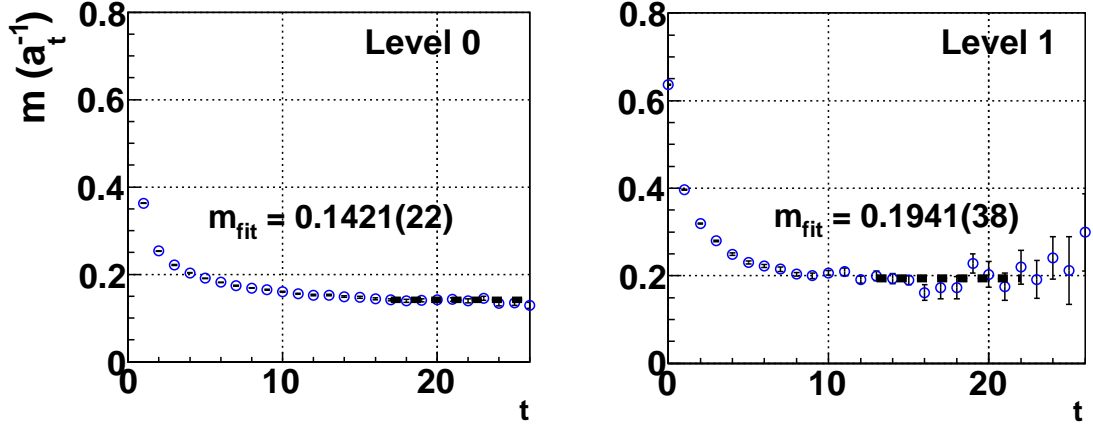}
  \caption{Effective masses corresponding to the diagonal elements of the rotated
$2\times 2$ correlator matrix involving a single-site $\rho$-meson operator and
an $I=1$ $\pi\pi$ operator in a $P$-wave with minimal relative momentum.  The $\rho$
operator dominates the lowest-lying level (left), while the $\pi\pi$ operator
dominates the first-excited state (right).  Although the mixing of the operators
is small, it is not negligible.  These results were obtained on 584 configurations
of the $24^3\times 128$ lattice with pion mass $m_\pi\sim 240$~MeV.
\label{fig:rhopipi}}
\end{figure}

Some initial results that incorporate multi-hadron operators are shown in
Fig.~\ref{fig:rhopipi}.  A $2\times 2$ correlation matrix was evaluated
involving a single-site $\rho$-meson operator and a total isospin $I=1$
$\pi\pi$ operator in a $P$-wave with minimal relative momentum. 
The stochastic LapH method enables very accurate
estimates of all elements (both diagonal and off-diagonal) of this
correlation matrix such that diagonalization can be done.  The effective
masses associated with the diagonalized correlator are shown in
Fig.~\ref{fig:rhopipi}. The $\rho$ operator dominates the lowest-lying level,
while the $\pi\pi$ operator dominates the first-excited state.  Although 
the mixing of the operators is small, it is not negligible.  This is 
certainly a warning about the dangers of extracting high-lying resonance
energies using only single-hadron operators.

\begin{figure}[t]
\begin{center}
  \includegraphics[width=4in,bb=0 15 567 306]{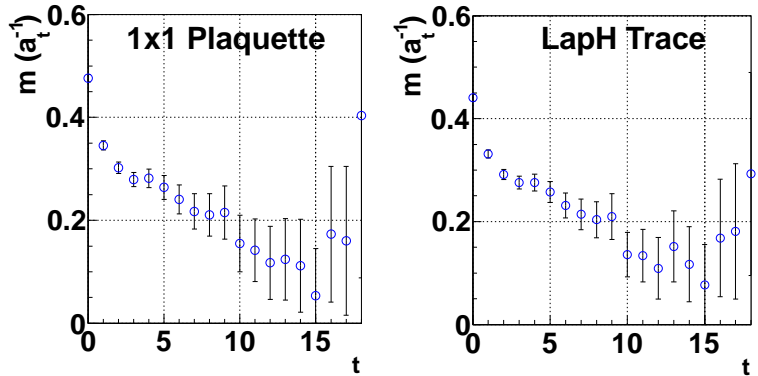}
\end{center}
  \caption{Comparison of the effective mass associated with the correlator
of the standard smeared plaquette glueball operator (left) with that of our
new glueball operator $G_\Delta$ defined in the text (right).  The similarity of the 
two results shows that the new glueball operator is just as useful as the
familiar smeared plaquette for studying the scalar glueball.  These effective
masses do not reach a plateau at the glueball mass since various $\pi\pi$
states and other multi-hadron states have smaller energies than the glueball mass.
These results were obtained using 584 configs of the $24^3$ ensemble for pion mass
$m_\pi\sim 240$~MeV. \label{fig:glueball}}
\end{figure}

Determining meson masses in the interesting scalar isoscalar sector
will ultimately involve including a scalar glueball operator, so we began
looking into the feasibility of such calculations.  LapH quark-field smearing
involves the covariant spatial Laplacian $\widetilde{\Delta}$.  The eigenvalues 
of the Laplacian
are invariant under rotations and gauge transformations so are appropriate
for a scalar glueball operator.  The lowest-lying eigenvalue was studied,
as well as other functions of the eigenvalues.  We found that any
combination of the low-lying eigenvalues worked equally well for
studying the scalar glueball.  In particular, the operator defined by
$G_\Delta(t) = 
-{\rm Tr}(\Theta(\sigma_s^2+\widetilde{\Delta})\widetilde{\Delta}(t))$
was used.  The effective mass associated with this operator is compared
to that of the smeared plaquette glueball operator in
Fig.~\ref{fig:glueball}.  The similarity of the results shows that $G_\Delta$
is just as useful as the familiar smeared plaquette for studying the scalar 
glueball. 

\begin{figure}[t]
  \includegraphics[width=5.8in,bb=0 15 567 155]{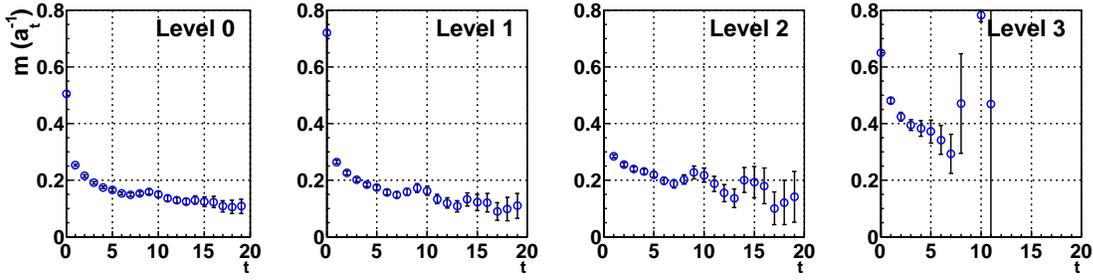}
  \caption{Effective masses corresponding to the diagonal elements of the rotated
$4\times 4$ correlator matrix involving a scalar isoscalar single-site 
quark-antiquark meson operator, the new glueball operator $G_\Delta$, and
two $I=0$ $\pi\pi$ operators in an $S$ wave (one with zero relative momentum and
the other with minimal nonzero relative momentum). Mixing of these operators is 
sizeable. More $\pi\pi$ operators must be included to reliably extract the glueball
mass.  These results were obtained on 100 configurations
of a $16^3\times 128$ lattice with pion mass $m_\pi\sim 390$~MeV.
\label{fig:isoscalar}}
\end{figure}

Effective masses corresponding to the diagonalized $4\times 4$ correlator 
matrix involving a scalar isoscalar single-site quark-antiquark meson operator, 
the new glueball operator $G_\Delta$, and two $I=0$ $\pi\pi$ operators in an 
$S$ wave (one with zero relative momentum and the other with minimal nonzero 
relative momentum) are shown in Fig.~\ref{fig:isoscalar}. Mixing of these 
operators is sizeable. More $\pi\pi$ operators must be included to reliably 
extract the glueball mass. 

\section{Conclusion}

A new method of stochastically estimating the low-lying effects of quark
propagation was proposed which allows accurate determinations of temporal
correlations of single-hadron and multi-hadron operators in lattice QCD.
The method enables accurate treatment of hadron correlators involving
quark propagation from all spatial sites on one time slice to all spatial
sites on another time slice.  Contributions involving quark lines 
originating at the sink time $t_F$ and terminating at the same sink time $t_F$
are easily handled, even for a large number of $t_F$ times.

The effectiveness of the method can be traced to two of its key features: the
use of noise dilution projectors that interlace in time and the use of
$Z_N$ noise in the subspace defined by the Laplacian Heaviside quark-field
smearing.  Introducing noise in the LapH subspace results in greatly
reduced variances in temporal correlations compared to methods that 
introduce noise on the entire lattice.  Although the number of Laplacian
eigenvectors needed to span the LapH subspace rises linearly with the
spatial volume, we found that the number of inversions of the Dirac matrix
needed for a target accuracy was remarkably insensitive to the lattice volume,
once a sufficient number of dilution projectors were introduced.

In addition to increased efficiency, the stochastic LapH method has other
advantages.  The method leads to complete factorization of hadron 
sources and sinks in temporal correlations, which greatly simplifies 
the logistics of evaluating correlation matrices involving large numbers
of operators.  Implementing the Wick contractions of the quark lines is
also straightforward.  Contributions from different Wick orderings 
within a class of quark-line diagrams differ only by permutations of the
noises at the source. 

The results presented here demonstrate that the stochastic LapH method is
useful for accurately estimating all of the temporal correlations needed 
for a full study of the QCD stationary-state energy spectrum, which we are 
currently pursuing.   A new glueball operator was tested and 
computing the mixing of this glueball operator with a quark-antiquark
operator and multiple two-pion operators was shown to be feasible.  Some of
our initial results showed warning signs about extracting high-lying
resonance energies using only single-hadron operators.

This work was supported by the U.S.~NSF
under awards PHY-0510020, PHY-0653315, PHY-0704171, PHY-0969863, and
PHY-0970137, and through TeraGrid/XSEDE resources provided by 
TACC and NICS under grant numbers TG-PHY100027 and TG-MCA075017.

\end{document}